\documentclass[11pt]{article}
\usepackage[utf8]{inputenc}
\usepackage{amssymb,amsmath}
\usepackage{graphicx,units,mathrsfs}
\usepackage{subfigure}
\usepackage[sort&compress,numbers]{natbib} 
\usepackage{hyperref,xcolor}
\usepackage[paperwidth=8.5in, paperheight=11in, portrait, top=1in, bottom=1.in, left=1.15in, right=1.15in]{geometry}
\usepackage{tikz}
\usetikzlibrary{calc,decorations.markings,positioning}

\newcommand\btd{\raise 2pt \hbox{$\hat\bigtriangledown$}\hskip 1.5pt}
\newcommand\bt{\raise 2pt \hbox{$\bigtriangledown$}\hskip 1.5pt}

\newtheorem{rmk}{Remark}[section]

\def\osp{{\mathfrak{osp}}}

\def\sl{{\mathfrak{sl}}}
\def\os{{\mathfrak{o}}}
\def\su{{\mathfrak{su}}}
\def\sp{{\mathfrak{sp}}}

\newcommand{\wt}[1]{\widetilde{#1}}
\newcommand{\ol}[1]{\overline{#1}}

\newcommand{\ket}[1]{|#1\rangle}

\def\qqquad{\qquad\quad}

\numberwithin{equation}{section}

\begin{document}
\title{\bf The dual pair $\big(U_q(\mathfrak{su}(1,1)),\mathfrak{o}_{q^{1/2}}(2n)\big)$, $q$-oscillators and Askey-Wilson algebras}
\author{
Luc Frappat \footnote{E-mail: luc.frappat@lapth.cnrs.fr} \\
\small~Laboratoire d'Annecy-le-Vieux de Physique Th\'eorique LAPTh, \\
\small~Univ. Grenoble Alpes, Univ. Savoie Mont Blanc, CNRS, F-74000 Annecy, France.\\[.9em]
Julien Gaboriaud \footnote{E-mail: gaboriaud@CRM.UMontreal.CA} \\
\small~Centre de Recherches Math\'ematiques, Universit\'e de Montr\'eal, \\
\small~P.O. Box 6128, Centre-ville Station, Montr\'eal (Qu\'ebec), H3C 3J7, Canada.\\[.9em]
Eric Ragoucy \footnote{E-mail: eric.ragoucy@lapth.cnrs.fr}\\
\small~Laboratoire d'Annecy-le-Vieux de Physique Th\'eorique LAPTh, \\
\small~Univ. Grenoble Alpes, Univ. Savoie Mont Blanc, CNRS, F-74000 Annecy, France.\\[.9em]
Luc Vinet \footnote{E-mail: vinet@CRM.UMontreal.CA} \\
\small~Centre de Recherches Math\'ematiques, Universit\'e de Montr\'eal, \\\small~P.O. Box 6128, Centre-ville Station, Montr\'eal (Qu\'ebec), H3C 3J7, Canada.
}
\date{}
\maketitle

\noindent{\bf Abstract:} The universal Askey--Wilson algebra $AW(3)$ can be obtained as the commutant of $U_q(\mathfrak{su}(1,1))$ in $U_q(\mathfrak{su}(1,1))^{\otimes3}$. We analyze the commutant of $\mathfrak{o}_{q^{1/2}}(2)\oplus\mathfrak{o}_{q^{1/2}}(2)\oplus\mathfrak{o}_{q^{1/2}}(2)$ in $q$-oscillator representations of $\mathfrak{o}_{q^{1/2}}(6)$ and show that it also realizes $AW(3)$. These two pictures of $AW(3)$ are shown to be dual in the sense of Howe; this is made clear by highlighting the role of the intermediate Casimir elements of each members of the dual pair $\big(U_q(\mathfrak{su}(1,1)),\mathfrak{o}_{q^{1/2}}(6)\big)$. We also generalize these results. A higher rank extension of the Askey--Wilson algebra denoted $AW(n)$ can be defined as the commutant of $U_q(\mathfrak{su}(1,1))$ in $U_q(\mathfrak{su}(1,1))^{\otimes n}$ and a dual description of $AW(n)$ as the commutant of $\mathfrak{o}_{q^{1/2}}(2)^{\oplus n}$ in $q$-oscillator representations of $\mathfrak{o}_{q^{1/2}}(2n)$ is offered by calling upon the dual pair $\big(U_q(\mathfrak{su}(1,1)),\mathfrak{o}_{q^{1/2}}(2n)\big)$.\\



\section{Introduction}
The Askey--Wilson algebra encodes the bispectrality properties of the eponym polynomials \cite{Zhedanov1991}. It is finding applications in various areas such as integrable models \cite{Baseilhac2005,Aneva2008,Vinet2008,Baseilhac2005a,Baseilhac2005b}, algebraic combinatorics \cite{Terwilliger2005,Terwilliger2011,Terwilliger2013,Terwilliger2018}, knot theory \cite{Bullock1999}, double affine Hecke algebras and representation theory \cite{Koornwinder2006,Koornwinder2008,Mazzocco2016}, etc. A universal extension is known to arise in an algebraic description of the Racah problem for $U_q(\sl_2)$ \cite{Granovskii1993a,Huang2017}. The goal of the present paper is to enlarge the fundamental understanding of this algebra by casting it in an alternate framework. We shall offer a picture of the universal Askey--Wilson algebra that is dual to the one which arises in the coupling of three representations of $U_q(\sl_2)$ and that will be reviewed in Section \ref{sec:AW3}. The presentation that is the object of this paper will be deemed dual, in the sense of Howe, to the $U_q(\sl_2)$ tensorial approach because it will rely on the complementary member $\os_{q^{1/2}}(2n)$ of the dual pair $\big(U_q(\mathfrak{su}(1,1)),\mathfrak{o}_{q^{1/2}}(2n)\big)$.

The concept of dual pairs has been introduced by Howe in \cite{Howe1989, Howe1989a} and has been since connected to numerous physical models (for a non-exhaustive list see \cite{Howe1987, Moshinsky1971, Jakobsen1977} and references therein). Let us recall the definition of the dual pairs in the context of Lie groups and Lie algebras \cite{Howe1987}:

\emph{Let $S$ be a Lie group and let $G$, $G'$ be a pair of subgroups of $S$. We say $(G,G')$ form a \emph{dual pair} of subgroups of $S$ if $G'$ is the full commutant of $G$ in $S$, and vice versa. The pair $(\mathfrak{g},\mathfrak{g}')$ of Lie algebras of $(G,G')$ are a dual pair in the Lie algebra $\mathfrak{s}$ of $S$.}

When each subgroup of the pair is reductive (that is, completely reducible), the pair is referred to as a \textit{reductive dual pair}. For the more technical details on the classification of these pairs, see for instance \cite{Howe1989, Howe1989a, Rowe2011, Rowe2012}. If one of the members of the pair is a compact group, the following decomposition holds:

\emph{Consider a Hilbert space $\mathscr{H}$ which supports representations of $S$. Then, the actions of $G$ and $G'$ on $\mathscr{H}$ commute, the reductive dual pair $(G,G')$ admits \emph{dual representations} on $\mathscr{H}$ and one obtains a multiplicity-free decomposition of the form:
\begin{align}\label{eq:pairing_modules}
 \mathscr{H}=\bigoplus_{\lambda} \Gamma^{(\lambda)}\otimes\Gamma\,'^{\,(\lambda)}
\end{align}
where $\Gamma$'s and $\Gamma\,'$'s are irreducible modules of $G$ and $G'$ respectively.} (Note that a similar decomposition occurs in the context of the Schur-Weyl duality.) In simpler words, the irreps of each member of the pair are matched together. By virtue of the exponential mapping correspondance between Lie algebras and Lie groups, a decomposition of the form \eqref{eq:pairing_modules} also holds for irreducible modules of the Lie algebras $(\mathfrak{g},\mathfrak{g}')$. We will be working at the algebra level in what follows.\\[.em]

An example of particular interest resides with the algebras $\sp(2)\simeq\su(1,1)$ and $\os(2n)$ which form a dual pair in $\sp(4n)$. This dual pair led to a novel interpretation of the Racah algebra $R(n)$ as the commutant of the $\os(2)^{\oplus n}$ algebra in oscillator representations of $\os(2n)$ \cite{Gaboriaud2018, Gaboriaud2018a}. To this end, models of both the $\su(1,1)^{\otimes 2n}$ and the $\os(2n)$ algebras were constructed in terms of $2n$ oscillators. Due to the decomposition \eqref{eq:pairing_modules}, the irreps could be paired. As expected by Schur's lemma, the pairing was expressed through the Casimirs which label the irreps. This was done as follows: 
The commutant of $\os(2)^{\oplus n}$ in $\os(2n)$ could be identified as the algebra generated by the quadratic Casimirs corresponding to $\os(2m)$ embeddings in $\os(2n)$ with $m=1,\dots,n$. In the oscillator model, all these quadratic Casimirs of $\os(2m)$ were seen to be affinely related to some $\su(1,1)$ Casimirs that arise from the recoupling of $2m$ copies of $\su(1,1)$. But, the Racah algebra $R(n)$ obtained as the commutant of $\su(1,1)$ in $U(\su(1,1))^{\otimes n}$ is precisely generated by these $\su(1,1)$ Casimirs occuring in the recoupling of $2m$ copies. Thus the two ``dual'' types of commutants give rise to the same Racah algebra. The pairing of the irreps \eqref{eq:pairing_modules} proves evident because the Casimirs of the two members of the pair are affinely related to each other, and it is this pairing that is fundamentally at the root of the dual descriptions of the Racah algebra $R(n)$.\\[.em]

The results on dual pairs that have been presented so far involve the so-called classical dual pairs. One may wonder what happens if $q$-deformations of the classical Lie algebras $\sl_2$ and $\os(2n)$ are considered, and if $q$-analogs of the dual pairs can be defined, while preserving a result analogous to \eqref{eq:pairing_modules} for the pairing of the irreps. 
Remarkably, a $q$-deformation of the pair $(\sl_2,\os(n))$ has been defined in \cite{Noumi1996}. The $q$-deformed dual pair $\big(U_q(\sl_2),\os_{q^{1/2}}(n)\big)$ will be used in an approach similar to the one employed for the Racah algebra to obtain dual pictures of the Askey--Wilson algebra. Note that we will actually restrict ourselves to the real form $U_q(\su(1,1))$ of $U_q(\sl_2)$ throughout this paper as this will allow to highlight more easily how the $q\to1$ limit connects with the results in \cite{Gaboriaud2018, Gaboriaud2018a}.\\[.em]

Here is the outline of the rest of the paper. In Section \ref{sec:AW3}, the (universal) Askey-Wilson algebra will be introduced along with its relation to $U_q(\su(1,1))$. The $q$-oscillator algebra will be defined in Section \ref{sec:qosc} and then used to build realizations of the $U_q(\su(1,1))$ and $\os_{q^{1/2}}(m)$ algebras. In Section \ref{sec:comm_AW3}, the Askey--Wilson algebra $AW(3)$ will be obtained as the ``dual'' commutant and this result will then be generalized to $AW(n)$ in Section \ref{sec:comm_AWn}. Concluding remarks and opening questions will complete the paper.

\section{A brief review of the Askey--Wilson algebra}\label{sec:AW3}

\subsection{The Askey--Wilson algebra $AW(3)$}
The Askey--Wilson algebra was first introduced by Zhedanov in \cite{Zhedanov1991}. It can be presented in terms of two generators $K_0$, $K_1$ obeying the $q$-commutation relations
\begin{align}\label{eq:ZhedanovAW}
 [K_0,K_1]_q=K_2,\qquad 
\begin{aligned}{}
 [K_1,K_2]_q&=bK_1+c_0K_0+d_0,\\
 [K_2,K_0]_q&=bK_0+c_1K_1+d_1,
\end{aligned}
\end{align}
where $b$, $c_0$, $c_1$, $d_0$, $d_1$ are (real) structure constants  and $[A,B]_q=qAB-q^{-1}BA$. Throughout the paper, we consider the case where $q$ is not a root of unity.

It is straightforward to reabsorb a few structure constants and to rescale the generators in order to arrive to the following $\mathbb{Z}_3$-symmetric presentation. Taking
\begin{align}
\begin{aligned}
 K_A&=-\frac{q^{2}-q^{-2}}{\sqrt{c_1}}K_0,\\
 K_B&=-\frac{q^{2}-q^{-2}}{\sqrt{c_0}}K_1,\\
 K_C&=-\frac{q^{2}-q^{-2}}{\sqrt{c_0\,c_1}}\left(K_2-\frac{b}{q-q^{-1}}\right),
\end{aligned}\qqquad
\begin{aligned}
 \alpha&=\frac{d_0}{c_0\sqrt{c_1}}(q+q^{-1})^{2}(q-q^{-1}),\\
 \beta&=\frac{d_1}{c_1\sqrt{c_0}}(q+q^{-1})^{2}(q-q^{-1}),\\
 \gamma&=\frac{b}{\sqrt{c_0\,c_1}}(q+q^{-1})^{2},
\end{aligned}
\end{align}
relations \eqref{eq:ZhedanovAW} are rewritten as
\begin{align}\label{eq:universalAW}
\begin{aligned}
 \frac{[K_A,K_B]_q}{q^{2}-q^{-2}}+K_C&=\frac{\gamma}{q+q^{-1}},\\
 \frac{[K_B,K_C]_q}{q^{2}-q^{-2}}+K_A&=\frac{\alpha}{q+q^{-1}},\\
 \frac{[K_C,K_A]_q}{q^{2}-q^{-2}}+K_B&=\frac{\beta}{q+q^{-1}}.
\end{aligned}
\end{align}
The universal Askey--Wilson algebra \cite{Terwilliger2011} is defined by the relations \eqref{eq:universalAW} with $\alpha$, $\beta$, $\gamma$ being central elements. The universal Askey--Wilson algebra is the one that will be referred to in the remainder of this paper.

\subsection{The $U_q(\su(1,1))$ algebra and its Racah problem}
We now review how the (universal) Askey--Wilson algebra appears in the context of the Racah problem of $U_q(\su(1,1))$.

The $U_q(\su(1,1))$ algebra has three generators, $J_\pm$ and $J_0$, obeying 
\begin{align}\label{eq:relations_uqsu11}
[ J_0 \,, J_\pm ] = \pm J_\pm, \qquad J_-J_+ - q^2 J_+J_- = q^{2J_0}\,[ 2J_0 ]_q. 
\end{align}
Here the notation $[x]_q$ stands for the $q$-number:
\begin{align}
 [x]_q=\frac{q^{x}-q^{-x}}{q-q^{-1}}.
\end{align}
This algebra can be endowed with a Hopf structure; in particular it posesses a coproduct which is an algebra morphism that defines an embedding of $U_q(\su(1,1))$ in $U_q(\su(1,1))\otimes U_q(\su(1,1))$:
\begin{align}\label{eq:coprod_uqsu11}
\Delta (J_0) = J_0 \otimes 1 + 1 \otimes J_0 , \qquad \Delta (J_\pm) = J_\pm \otimes q^{2J_0} + 1 \otimes J_\pm .
\end{align}
The Casimir operator $C$ of $U_q(\su(1,1))$ has the following expression
\begin{align}\label{eq:casimirsu}
\begin{aligned}
C &= J_+J_- q^{-2J_0+1} + (J_0)_{q^{2}}(1-J_0)_{q^{2}}
,
\end{aligned}
\end{align}
where the notation $(x)_q$ stands for another type of $q$-number:
\begin{align}
 (x)_q=\frac{1-q^{x}}{1-q}.
\end{align}
\begin{rmk} In the limit $q \to 1$, one recovers the usual $\mathfrak{su}(1,1)$ Lie algebra. The Casimir of $U_q(\su(1,1))$ that is being used has been shifted by a constant with respect to the more conventional Casimir
\begin{align}
 C=J_+J_- q^{-2J_0+1} - \frac{q}{(1-q^2)^2} \, \big( q^{2J_0-1} + q^{-2J_0+1} \big)
\end{align}
so as to make sure the $q\to1$ limit is non-singular and yields the usual $\su(1,1)$ Casimir $C=J_+J_--{J_0}^{2}+J_0$. Moreover, the standard presentation of $U_q(\mathfrak{su}(1,1))$ \cite{Klimyk1997} is recovered from \eqref{eq:relations_uqsu11} if one considers instead the generators $\wt J_0=J_0$,~ $\wt J_+ = J_+q^{-J_0}$ and $\wt J_- = q^{-J_0}J_-$, which satisfy the commutation relations $[\wt J_0 \,, \wt J_\pm] = \pm \wt J_\pm$, $[\wt J_- \,, \wt J_+] = [ 2\wt J_0]_q$.
\end{rmk}

Let us now consider the addition of three irreducible representations of $U_q(\su(1,1))$. Associated to each of those copies of $U_q(\su(1,1))$ are the Casimirs $C^{(i)}$, $i=1,2,3$. The coassociativity of the coproduct ensures that the following two ways to pair the representations
\begin{align}\label{eq:embeddings3}
 (1\oplus2)\oplus3 \simeq 1\oplus(2\oplus3)
\end{align}
are equivalent. 

In addition to the initial Casimirs $C^{(i)}$, there are the intermediate Casimirs associated to the different embeddings shown above in \eqref{eq:embeddings3}, $C^{(12)}=\Delta(C)\otimes1$ and $C^{(23)}=1\otimes\Delta(C)$, as well as a total Casimir operator, $C^{(123)}=\Delta^{(2)}(C)$. Here, we use the notation $\Delta^{(N)}=(1^{\otimes (N-1)}\otimes\Delta)\Delta^{(N-1)}$, with $\Delta^{(0)}=1$. The Racah problem of $U_q(\su(1,1))$ is to find the overlap between the two different bases corresponding to \eqref{eq:embeddings3}, i.e. the one which diagonalizes $C^{(12)}$ and the other which diagonalizes $C^{(23)}$.

The connection with the Askey--Wilson algebra follows from the fact that the intermediate Casimirs of $U_q(\su(1,1))$ realize it. In other words, the Askey--Wilson algebra can be described as the commutant of $U_q(\mathfrak{su}(1,1))$ in $U_q(\mathfrak{su}(1,1))^{\otimes3}$. Indeed, performing the affine transformation
\begin{align}\label{eq:realiz_AW3_cas}
 C_{\{\bullet\}}=-q(q-q^{-1})^{2}C^{(\bullet)}+(q+q^{-1}),
\end{align}
one checks that relations \eqref{eq:universalAW} are verified for 
\begin{align}\label{eq:realiz_AW3_cnst}
\begin{aligned}
 K_A&=C_{\{1,2\}},\\
 K_B&=C_{\{2,3\}},
\end{aligned}\qqquad
\begin{aligned}
 \alpha&=C_{\{2\}}C_{\{3\}}+C_{\{1\}}C_{\{1,2,3\}},\\
 \beta&=C_{\{1\}}C_{\{3\}}+C_{\{2\}}C_{\{1,2,3\}},\\
 \gamma&=C_{\{1\}}C_{\{2\}}+C_{\{3\}}C_{\{1,2,3\}}.
\end{aligned}
\end{align}
It is worth noting that instead of looking at the $q$-commutation relations of the intermediate Casimirs, one could have instead presented the relations using commutation relations. With $Q_1=C^{(12)}$, $Q_2=C^{(23)}$, one obtains the following relations:
\begin{align}\label{eq:qRacah}
\begin{aligned}{}
 [Q_1,Q_2] &= Q_3,\\
 [Q_2,Q_3] &= r Q_2 Q_1 Q_2 + \xi_1 \{ Q_1,Q_2 \} + \xi_2 Q_2^2 + \xi_3 Q_2 + \xi_5, \\
 [Q_3,Q_1] &= r Q_1 Q_2 Q_1 + \xi_1 Q_1^2 + \xi_2 \{ Q_1,Q_2 \} + \xi_3 Q_1 + \xi_7, 
\end{aligned}
\end{align}
where the parameters $r$ and $\xi_i$ take the following values:
\begin{align}\label{eq:param_qRacah}
 r&=-(q-q^{-1})^2, \qqquad \xi_1 = \xi_2 = (1+q^{-2}),\nonumber\\
 \xi_3&=-r \big(C^{(1)}C^{(3)}+C^{(2)}C^{(123)}\big) - \xi_1 \, \big(C^{(1)}+C^{(2)}+C^{(3)}+C^{(123)}\big),\\
 \xi_5&=\xi_1 \big(C^{(2)}-C^{(3)}\big)\big(C^{(1)}-C^{(123)}\big), \qqquad \xi_7 = \xi_1 \big(C^{(2)}-C^{(1)}\big)\big(C^{(3)}-C^{(123)}\big). \nonumber 
\end{align}
The $q\to1$ limit of this algebra immediately leads to the (classical) Racah algebra.

\section{$q$-oscillator realization of the dual pair $\big(U_q(\mathfrak{su}(1,1)),\mathfrak{o}_{q^{1/2}}(2n)\big)$}\label{sec:qosc}

\subsection{The $\mathfrak{o}_{q^{1/2}}(N)$ algebra}
The non-standard $q$-deformation $\mathfrak{o}_{q^{1/2}}(N)$ of $\mathfrak{o}(N)$, often denoted $U'_{q^{1/2}}(\mathfrak{so}_N)$ in the litterature \cite{Gavrilik2000,Klimyk2001,Klimyk2002,Iorgov2005}, can be defined as the associative unital algebra with generators $L_{i,i+1}$ ($i=1,\dots,N-1$) obeying the relations
\begin{subequations}\label{eq:soqm}
\begin{align}
& L_{i-1,i}\,L_{i,i+1}^2 - (q^{1/2}+q^{-1/2}) L_{i,i+1}\,L_{i-1,i}\,L_{i,i+1} + L_{i,i+1}^2\,L_{i-1,i} = -L_{i-1,i},\\
& L_{i,i+1}\,L_{i-1,i}^2 - (q^{1/2}+q^{-1/2}) L_{i-1,i}\,L_{i,i+1}\,L_{i-1,i} + L_{i-1,i}^2\,L_{i,i+1} = -L_{i,i+1},\\
& [ L_{i,i+1}, L_{j,j+1} ] = 0 \quad \text{for} \quad |i-j| > 1 . \label{eq:soqm3}
\end{align}
\end{subequations}
This algebra posesses numerous properties of interest, among which: $\mathfrak{o}_q(N)$ can be viewed as a $q$-analogue of the symmetric space based on the pair $(\mathfrak{gl}(N),\mathfrak{o}(N))$ \cite{Noumi1996a}, it is a coideal subalgebra of $U_q(\mathfrak{sl}(N))$ \cite{Noumi1996a} and appears in various areas of mathematical physics \cite{Klimyk2002}.

The quadratic Casimir of $\mathfrak{o}_{q^{1/2}}(2n)$ is given in \cite{Noumi1996,Gavrilik2000}. We shall need the following elements:
\begin{align}\label{eq:oq_ij}
L_{ik}^\pm = [L_{ij}^{\pm} \,, L_{jk}^{\pm}]_{q^{\pm 1/4}}=q^{\pm1/4}L_{ij}^{\pm}L_{jk}^{\pm}-q^{\mp1/4}L_{jk}^{\pm}L_{ij}^{\pm},\qquad \text{for any}\quad i<j<k
\end{align}
with $L_{i,i+1}^\pm=L_{i,i+1}$ by definition. The quadratic Casimir operator of the algebra $\mathfrak{o}_{q^{1/2}}(2n)$ then has the following expression: 
\begin{align}\label{eq:oqn_cas}
\Lambda^{[2n]}&=\sum_{1\leq i<j\leq 2n} q^{\frac{-2n+i+j-1}{2}}L_{ij}^{+}L_{ij}^{-}.
\end{align}

\subsection{The $q$-oscillator algebra}
The $q$-oscillator algebra $\mathcal{A}_q(N)$ \cite{Klimyk1997} is the unital associative algebra over $\mathbb{C}$ generated by $N$ independent sets of $q$-oscillators $\{A_i^\pm$, $A_i^0\}$ verifying
\begin{align}\label{eq:AqN_relations}
[A_i^0, A_i^\pm] = \pm A_i^\pm, \qquad [A_i^-, A_i^+] = q^{A_i^0}, \qquad A_i^- A_i^+ - q A_i^+ A_i^- = 1,\qquad i=1,\dots,N,
\end{align}
and such that the commutators between elements with distinct indices $i$ are equal to zero.
The last two relations lead to:
\begin{align}
A_i^+ A_i^- = \frac{1-q^{A_i^0}}{1-q} =(A_i^0)_q.
\end{align}
The $q$-oscillator algebra $\mathcal{A}_q(N)$ admits an irreducible representation bounded from below with orthonormal basis vectors $|n_1,\cdots,n_N\rangle = |n_1\rangle \otimes \cdots \otimes |n_N\rangle$ and with the operators $A_i$ acting on the $i$'th factor of the states according to:
\begin{align}\label{eq:AqN_actions}
A_i^0 |n_i\rangle = n_i |n_i\rangle , \qquad A_i^+ |n_i\rangle = \sqrt{\frac{1-q^{n_i+1}}{1-q}} |n_i+1\rangle , \qquad A_i^- |n_i\rangle = \sqrt{\frac{1-q^{n_i}}{1-q}} |n_i-1\rangle .
\end{align}
These commuting $q$-oscillators can now be used to realize the algebras considered previously.

\subsection{Dual realizations of the $\mathfrak{o}_{q^{1/2}}(2n)$ and $U_q(\su(1,1))$ algebras}
The algebras $U_q(\mathfrak{su}(1,1))$ and $\mathfrak{o}_{q^{1/2}}(2n)$ can be realized in terms of $q$-oscillators and shown to have commuting actions on the Hilbert space of $q$-oscillators. 

To that end, let us first consider $2n$ copies of the $q$-deformation of the usual metaplectic representation of $\mathfrak{su}(1,1)$, which is realized with $2n$ $q$-oscillators by taking
\begin{align}\label{eq:metaplectic_map}
 \mathscr{J}_0^{i}=\frac{1}{2}\left(A^{0}_i+\frac{1}{2}\right),\qquad \mathscr{J}_\pm^{i}=\frac{1}{[2]_{q^{1/2}}}(A^{\pm}_i)^{2},\qqquad i=1,\dots,2n.
\end{align}
Owing to the fact that each set of $\mathscr{J}_0^{i}$, $\mathscr{J}_\pm^{i}$ acts only on the $i$'th oscillator and obeys the relations \eqref{eq:relations_uqsu11}, \eqref{eq:metaplectic_map} hence gives a realization of $U_q(\mathfrak{su}(1,1))^{\otimes 2n}$. It is then straightforward to embed $U_q(\mathfrak{su}(1,1))$ inside $\mathcal{A}_q(2n)$ by repeatedly making use of the coproduct \eqref{eq:coprod_uqsu11}:
\begin{align}
\begin{aligned}
 \mathscr{J}_0^{(2n)}&=\Delta^{(2n-1)}(\mathscr{J}_0)=\frac{1}{2}\sum_{i=1}^{2n}\left(A_i^{0}+\frac{1}{2}\right),\\
 \mathscr{J}_\pm^{(2n)}&=\Delta^{(2n-1)}(\mathscr{J}_\pm)=\frac{1}{[2]_{q^{1/2}}}\sum_{i=1}^{2n}\left((A_i^{\pm})^{2}\prod_{j=i+1}^{2n}q^{A_j^{0}+\frac{1}{2}}\right).
\end{aligned}
\end{align}
The algebra $\mathfrak{o}_{q^{1/2}}(2n)$ can also be realized in terms of $2n$ $q$-oscillators. The $2n-1$ generators take the form
\begin{align}\label{eq:Lscript_iip1}
 \mathscr{L}_{i,i+1}= q^{-\frac{1}{2}(A_i^0+\frac{1}{2})} \big( q^{\frac{1}{4}}A_i^+A_{i+1}^- - q^{-\frac{1}{4}}A_i^-A_{i+1}^+ \big),\qqquad i=1,\dots,2n-1.
\end{align}
A direct calculation shows that these $\mathscr{L}_{i,i+1}$'s verify the relations \eqref{eq:soqm}. All the other $\mathscr{L}_{ij}^{\pm}$'s can be obtained by \eqref{eq:oq_ij}.
\begin{rmk} In this particular realization, the following $q$-analogs of the angular momenta relation $M_{12}M_{34}+M_{13}M_{42}+M_{14}M_{23}=0$ in \cite{Feigin2015} hold. For $i<j<k<\ell$, one has:
\begin{align}
\begin{aligned}
 q^{-1/2}\mathscr{L}_{ij}^{+}\mathscr{L}_{k\ell}^{+}-\mathscr{L}_{ik}^{+}\mathscr{L}_{j\ell}^{+}+q^{+1/2}\mathscr{L}_{i\ell}^{+}\mathscr{L}_{jk}^{+}&=0,\\
 q^{+1/2}\mathscr{L}_{ij}^{-}\mathscr{L}_{k\ell}^{-}-\mathscr{L}_{ik}^{-}\mathscr{L}_{j\ell}^{-}+q^{-1/2}\mathscr{L}_{i\ell}^{-}\mathscr{L}_{jk}^{-}&=0.
\end{aligned}
\end{align}
\end{rmk}
It is easy to check that $[\mathscr{J}_0^{(2)}, \mathscr{L}_{12}]=[\mathscr{J}_\pm^{(2)}, \mathscr{L}_{12}]=0$. A straightforward induction argument using the coproduct \eqref{eq:coprod_uqsu11} and the form of the expression \eqref{eq:Lscript_iip1} leads to
\begin{align}\label{eq:commuting_actions}
 [\mathscr{J}_0^{(2n)}, \mathscr{L}_{i,i+1}]=[\mathscr{J}_\pm^{(2n)}, \mathscr{L}_{i,i+1}]=0,\qqquad i=1,\dots,2n-1.
\end{align}
In other words, $U_q(\mathfrak{su}(1,1))$ and $\mathfrak{o}_{q^{1/2}}(2n)$ have commuting actions on the Hilbert space of $2n$ $q$-oscillators. 

This feature precisely illustrates the Howe duality operating in this context and will be the key to obtaining the Askey--Wilson algebra of arbitrary rank as a ``dual'' commutant.

\section{The Askey--Wilson algebra $AW(3)$ as a ``dual'' commutant}\label{sec:comm_AW3}

\subsection{The commutant of $\mathfrak{o}_{q^{1/2}}(2)^{\oplus 3}$ in the $q$-oscillator realization of $\mathfrak{o}_{q^{1/2}}(6)$ and the Askey--Wilson algebra $AW(3)$}
We now look for the commutant of the $\mathfrak{o}_{q^{1/2}}(2)\oplus\mathfrak{o}_{q^{1/2}}(2)\oplus\mathfrak{o}_{q^{1/2}}(2)$ subalgebra generated by $\{\mathscr{L}_{12},\mathscr{L}_{34},\mathscr{L}_{56}\}$ in the $q$-oscillator realization of $\mathfrak{o}_{q^{1/2}}(6)$. From the expressions of the quadratic Casimirs \eqref{eq:oqn_cas} it is easy to identify the following $6$ independent elements:
\begin{align}
\begin{aligned}
 \Lambda_{1}&={\mathscr{L}_{12}}^{2},\\ \\
 \Lambda_{2}&={\mathscr{L}_{34}}^{2},\\ \\
 \Lambda_{3}&={\mathscr{L}_{56}}^{2},
\end{aligned}\qqquad
\begin{aligned}
 \Lambda_{12}&=q^{-1}{\mathscr{L}_{12}}^{2}+{\mathscr{L}_{23}^{+}}{\mathscr{L}_{23}^{-}}+q{\mathscr{L}_{34}}^{2}\\
 &\quad+q^{-1/2}{\mathscr{L}_{13}^{+}}{\mathscr{L}_{13}^{-}}+q^{1/2}{\mathscr{L}_{24}^{+}}{\mathscr{L}_{24}^{-}}+{\mathscr{L}_{14}^{+}}{\mathscr{L}_{14}^{-}},\\
 \Lambda_{23}&=q^{-1}{\mathscr{L}_{34}}^{2}+{\mathscr{L}_{45}^{+}}{\mathscr{L}_{45}^{-}}+q{\mathscr{L}_{56}}^{2}\\
 &\quad+q^{-1/2}{\mathscr{L}_{35}^{+}}{\mathscr{L}_{35}^{-}}+q^{1/2}{\mathscr{L}_{46}^{+}}{\mathscr{L}_{46}^{-}}+{\mathscr{L}_{36}^{+}}{\mathscr{L}_{36}^{-}},\\
 \Lambda_{13}&=q^{-1}{\mathscr{L}_{12}}^{2}+{\mathscr{L}_{25}^{+}}{\mathscr{L}_{25}^{-}}+q{\mathscr{L}_{56}}^{2}\\
 &\quad+q^{-1/2}{\mathscr{L}_{15}^{+}}{\mathscr{L}_{15}^{-}}+q^{1/2}{\mathscr{L}_{26}^{+}}{\mathscr{L}_{26}^{-}}+{\mathscr{L}_{16}^{+}}{\mathscr{L}_{16}^{-}}.
\end{aligned}
\end{align}
Instead of using $\Lambda_{13}$, one could alternatively take the element $\Lambda_{123}$ which is a linear combination of the other $\Lambda_{\bullet}$'s above
\begin{align}
 \Lambda_{123}=q^{-1}\Lambda_{12}+\Lambda_{13}+q\Lambda_{23}-(q^{-1}\Lambda_{1}+\Lambda_{2}+q\Lambda_{3})
\end{align}
and corresponds to the quadratic Casimir of $\mathfrak{o}_{q^{1/2}}(6)$ itself.

The algebraic relations obeyed by these $\Lambda_{\bullet}$'s correspond to those of the Askey--Wilson algebra. First make the affine transformation
\begin{align}
\begin{aligned}
 \widetilde{\Lambda}_{i}&=[2]_{q}-\left(\frac{q^{1/2}-q^{-1/2}}{1+q}\right)^{2}(\Lambda_{i}+1),\\
 \widetilde{\Lambda}_{ij}&=[2]_{q}-\left(\frac{q^{1/2}-q^{-1/2}}{1+q}\right)^{2}\Lambda_{ij},\\
 \widetilde{\Lambda}_{123}&=[2]_{q}-\left(\frac{q^{1/2}-q^{-1/2}}{1+q}\right)^{2}(\Lambda_{123}-[3]_{q^{1/2}}),
\end{aligned}
\end{align}
then take $K_A$ and $K_B$ to be
\begin{align}
\begin{aligned}
 K_A=\widetilde{\Lambda}_{12},\qquad K_B=\widetilde{\Lambda}_{23}.
\end{aligned}
\end{align}
A straightforward calculation shows that $K_A$ and $K_B$ obey the relations \eqref{eq:universalAW}, with structure constants $\alpha$, $\beta$, $\gamma$ expressible in terms of $\Lambda_1$, $\Lambda_2$, $\Lambda_3$ and $\Lambda_{123}$.
\begin{align}
\begin{aligned}
 \alpha&=\widetilde{\Lambda}_{2}\widetilde{\Lambda}_{3}+\widetilde{\Lambda}_{1}\widetilde{\Lambda}_{123},\\
 \beta&=\widetilde{\Lambda}_{3}\widetilde{\Lambda}_{1}+\widetilde{\Lambda}_{2}\widetilde{\Lambda}_{123},\\
 \gamma&=\widetilde{\Lambda}_{1}\widetilde{\Lambda}_{2}+\widetilde{\Lambda}_{3}\widetilde{\Lambda}_{123}.
\end{aligned}
\end{align}

\subsection{The $U_q(\mathfrak{su}(1,1))$ and $\mathfrak{o}_{q^{1/2}}(6)$ descriptions of $AW(3)$ and Howe duality}
In Section \ref{sec:AW3} the Askey--Wilson algebra was described as the commutant of $U_q(\mathfrak{su}(1,1))$ in $U_q(\mathfrak{su}(1,1))^{\otimes3}$. That the AW algebra could also be described as the commutant of $\mathfrak{o}_{q^{1/2}}(2)^{\oplus3}$ in $q$-oscillator representations of $\mathfrak{o}_{q^{1/2}}(6)$ is not a coincidence. We now make explicit the connection between these two approaches using Howe duality.

In \cite{Noumi1996} it was shown that $U_q(\mathfrak{su}(1,1))$ and $\mathfrak{o}_{q^{1/2}}(2m)$ are a dual pair in the sense of Howe. It follows from this that these two algebras have commuting actions on the Hilbert space of $2m$ $q$-oscillators (see \eqref{eq:commuting_actions}). This implies that their irreducible representations can be paired through the eigenvalues of the Casimirs that label them. We shall now indicate how the pairing occurs.

\medskip

Take $6$ $q$-oscillators, realizing $6$ copies of $U_q(\mathfrak{su}(1,1))$. We first couple them pairwise and label each couple by $\ol{\imath}\equiv(2i-1,2i)$ in order to obtain an embeddings of $U_q(\mathfrak{su}(1,1))^{\otimes3}$ in $U_q(\mathfrak{su}(1,1))^{\otimes6}$:
\begin{align}\label{eq:paired_uqsu11}
 \mathscr{J}_0^{\ol{\imath}}=\frac{1}{2}\left(A^{0}_{2i-1}+A^{0}_{2i}+1\right),\qquad \mathscr{J}_\pm^{\ol{\imath}}=\frac{1}{[2]_{q^{1/2}}}\left((A^{\pm}_{2i-1})^{2}q^{A^{0}_{2i}+\frac{1}{2}}+(A^{\pm}_{2i})^{2}\right),\qqquad i=1,\dots,3.
\end{align}
To each $\ol{\imath}$'th copy of $U_q(\mathfrak{su}(1,1))$ corresponds a Casimir $C^{\ol{\imath}}$ given by \eqref{eq:casimirsu}.

Three additional embeddings of $U_q(\mathfrak{su}(1,1))$ can be realized by repeatedly making use of the coproduct \eqref{eq:coprod_uqsu11}
\begin{align}
\begin{aligned}{}
 \mathscr{J}_0^{\ol{1}\,\ol{2}}&=\mathscr{J}_0^{\ol{1}}+\mathscr{J}_0^{\ol{2}},\\
 \mathscr{J}_0^{\ol{2}\,\ol{3}}&=\mathscr{J}_0^{\ol{2}}+\mathscr{J}_0^{\ol{3}},\\
 \mathscr{J}_0^{\ol{1}\,\ol{2}\,\ol{3}}&=\mathscr{J}_0^{\ol{1}}+\mathscr{J}_0^{\ol{2}}+\mathscr{J}_0^{\ol{3}},
\end{aligned}\qqquad
\begin{aligned}{}
 \mathscr{J}_\pm^{\ol{1}\,\ol{2}}&=\mathscr{J}_\pm^{\ol{1}}q^{2\mathscr{J}_0^{\ol{2}}}+\mathscr{J}_\pm^{\ol{2}},\\
 \mathscr{J}_\pm^{\ol{2}\,\ol{3}}&=\mathscr{J}_\pm^{\ol{2}}q^{2\mathscr{J}_0^{\ol{3}}}+\mathscr{J}_\pm^{\ol{3}},\\
 \mathscr{J}_\pm^{\ol{1}\,\ol{2}\,\ol{3}}&=\mathscr{J}_\pm^{\ol{1}}q^{2\mathscr{J}_0^{\ol{2}}}q^{2\mathscr{J}_0^{\ol{3}}}+\mathscr{J}_\pm^{\ol{2}}q^{2\mathscr{J}_0^{\ol{3}}}+\mathscr{J}_\pm^{\ol{3}},
\end{aligned}
\end{align}
and the respective Casimirs associated to each of these embeddings, $C^{\ol{1}\,\ol{2}}$, $C^{\ol{2}\,\ol{3}}$, $C^{\ol{1}\,\ol{2}\,\ol{3}}$ can be obtained from \eqref{eq:casimirsu}.

Schematically, these successive embeddings can be thought of as:
\begin{center}
\begin{tikzpicture}
\draw[fill] (-.25,0) circle [radius=0.08];
\draw[fill] (.25,0) circle [radius=0.08];
\node [above, black] at (-.25,.0) {$1$};
\node [above, black] at (.25,.0) {$2$};
\node [below, black] at (0,-.25) {$\ol{1}$};
\draw[postaction={decorate,decoration={markings,
      mark=at position 0.8 with {\node (a1) {};}}}]
      (0,.1) ellipse (0.6 and 0.4);

\draw[fill] (1,0) circle [radius=0.08];
\draw[fill] (1.5,0) circle [radius=0.08];
\node [above, black] at (1,.0) {$3$};
\node [above, black] at (1.5,.0) {$4$};
\node [below, black] at (1.25,-.25) {$\ol{2}$};
\draw[postaction={decorate,decoration={markings,
      mark=at position 0.8 with {\node (a2) {};}}}]
      (1.25,.1) ellipse (0.6 and 0.4);

\draw[fill] (2.25,0) circle [radius=0.08];
\draw[fill] (2.75,0) circle [radius=0.08];
\node [above, black] at (2.25,.0) {$5$};
\node [above, black] at (2.75,.0) {$6$};
\node [below, black] at (2.5,-.25) {$\ol{3}$};
\draw[postaction={decorate,decoration={markings,
      mark=at position 0.8 with {\node (a3) {};}}}]
      (2.5,.1) ellipse (0.6 and 0.4);

\draw(-.25,-1.05) arc (240:300:.5);
\node [below, black] at (0.,-1.) {$C^{\ol{1}}$};
\draw(1,-1.05) arc (240:300:.5);
\node [below, black] at (1.25,-1.) {$C^{\ol{2}}$};
\draw(2.25,-1.05) arc (240:300:.5);
\node [below, black] at (2.5,-1.) {$C^{\ol{3}}$};
\draw(-.25,-1.6) arc (240:300:1.75);
\draw(1,-1.6) arc (240:300:1.75);
\node [below, black] at (0.7,-1.75) {$C^{\ol{1}\,\ol{2}}$};
\node [below, black] at (1.95,-1.75) {$C^{\ol{2}\,\ol{3}}$};
\draw(-.25,-2.2) arc (240:300:3);
\node [below, black] at (1.25,-2.5) {$C^{\ol{1}\,\ol{2}\,\ol{3}}$};
\node [below, black] at (6.5,.5) {$U_q(\su(1,1))^{\otimes6}$};
\node [below, black] at (6.5,-.25) {$\bigcup$};
\node [below, black] at (6.5,-1) {$U_q(\su(1,1))^{\otimes3}$};
\node [below, black] at (6.5,-1.75) {$\bigcup$};
\node [below, black] at (6.5,-2.5) {$U_q(\su(1,1))$};
\end{tikzpicture}
\end{center}
Upon looking at the explicit expressions of these $C^{\ol{\imath}}$, $C^{\ol{\imath}\,\ol{\jmath}}$, $C^{\ol{1}\,\ol{2}\,\ol{3}}$ in terms of the $q$-oscillators, one finds that
\begin{align}\label{eq:duality_casimirs_3}
\begin{aligned}
 C^{\ol{\imath}}&=\frac{1}{(1+q)^{2}}\left(\Lambda_i+1\right),\\
 C^{\ol{\imath}\,\ol{\jmath}}&=\frac{1}{(1+q)^{2}}\left(\Lambda_{ij}\right),\qquad\quad \text{for consecutive $ij$'s}\\
 C^{\ol{1}\,\ol{2}\,\ol{3}}&=\frac{1}{(1+q)^{2}}\left(\Lambda_{123}-[3]_{q^{1/2}}\right).
\end{aligned}
\end{align}
Those expressions show how the intermediate Casimirs of $U_q(\su(1,1))$ and those of $\os_{q^{1/2}}(6)$ are affinely related. Let us emphasize that owing to the Howe duality between $U_q(\su(1,1))$ and $\os_{q^{1/2}}(6)$, the multiplicity-free decomposition of the form \eqref{eq:pairing_modules} takes place; the relations \eqref{eq:duality_casimirs_3} make this explicit keeping in mind the Schur's lemma.

Moreover, this pairing of the Casimirs is precisely what is behind the fact that the Askey--Wilson algebra, usually obtained from intermediate $U_q(\su(1,1))$ Casimirs, is expressible as the commutant of the $\os_{q^{1/2}}(2)^{\oplus3}$ algebra in $q$-oscillator representations of $\os_{q^{1/2}}(6)$. The duality of the two pictures is thus expressed in \eqref{eq:duality_casimirs_3}.

\section{The case for general $n$ and the algebra $AW(n)$}\label{sec:comm_AWn}

\subsection{Towards the higher rank Askey--Wilson algebra}
Let us first introduce the notation $[i;j]$ for sets of consecutive integers:
\begin{align}
 [i;j]=\begin{cases}
             \quad\{i,i+1,\dots,j\} \qqquad& j>i,\\
             \quad\{i\} & j=i,\\
             \quad\emptyset & j<i.
            \end{cases}
\end{align}
In \cite{DeBie2017a}, a higher rank extension of the Racah algebra $R(n)$ was realized as the algebra of the intermediate Casimir elements in $U(\su(1,1))$ associated to embeddings (labelled by $A\subset[1;n]$) of $\su(1,1)$ in its $n$-fold tensor product. A generating set for $R(n)$ is given by the intermediate Casimir operators related to consecutive tensor product space embeddings $[i;j]$, $1\leq i\leq j\leq n$.

A similar story is emerging for the Askey--Wilson algebra. The higher rank Askey--Wilson algebra $AW(n)$ has been defined tensorially as the algebra of the intermediate Casimir elements $C_A$ of $U_q(\su(1,1))$ associated to embeddings (labelled by $A\subset[1;n]$) of $U_q(\su(1,1))$ in its $n$-fold tensor product. A generating set of $AW(n)$ is given by all $C_{[i;j]}$'s, with $1\leq i\leq j\leq n$. These $C_{[i;j]}$'s are obtained from the repeated action of the coproduct on the $U_q(\su(1,1))$ Casimir elements.

The algebraic relations of $AW(4)$ are given in \cite{Post2017}. The full set of relations of $AW(n)$ is not known, however a large subset of those relations has been presented in \cite{DeBie2018}. Nevertheless, we here advance the understanding of these algebraic structures by establishing the dual connection between these intermediate Casimir $C_A$ in $U_q(\su(1,1))^{\otimes n}$ and the generators of the commutant of a subalgebra of $\os_{q^{1/2}}(2n)$.

\subsection{The Howe duality in the $AW(n)$ case}
We now proceed with this analysis of the higher rank case and look for the commutant of $\mathfrak{o}_{q^{1/2}}(2)^{\oplus n}$ in $\mathfrak{o}_{q^{1/2}}(2n)$. The algebra $\mathfrak{o}_{q^{1/2}}(2)^{\oplus n}$ is generated by the set $\{\mathscr{L}_{12},\,\dots,\,\mathscr{L}_{2n-1,2n}\}$. 

In view of the quadratic Casimirs \eqref{eq:oqn_cas}, we examine the following $\binom{n+1}{2}$ elements:
\begin{align}\label{eq:set_comm_n}
\begin{aligned}
 \Lambda_{i}&=({\mathscr{L}_{2i-1,2i}})^{2},\\
 \Lambda_{ij}&=q^{-1}{\mathscr{L}_{2i-1,2i}}^{2}+{\mathscr{L}_{2i,2j-1}^{+}}{\mathscr{L}_{2i,2j-1}^{-}}+q{\mathscr{L}_{2j-1,2j}}^{2}\\
 &\quad+q^{-1/2}{\mathscr{L}_{2i-1,2j-1}^{+}}{\mathscr{L}_{2i-1,2j-1}^{-}}+q^{1/2}{\mathscr{L}_{2i,2j}^{+}}{\mathscr{L}_{2i,2j}^{-}}+{\mathscr{L}_{2i-1,2j}^{+}}{\mathscr{L}_{2i-1,2j}^{-}}, 
\end{aligned}\qquad 1\leq i<j\leq n.
\end{align}
These elements all commute with $\mathfrak{o}_{q^{1/2}}(2)^{\oplus n}$:
\begin{align}
 [\Lambda_{i},\mathscr{L}_{2k-1,2k}]=[\Lambda_{ij},\mathscr{L}_{2k-1,2k}]=0,\qquad i<j,\quad i,j,k=1,\dots,n,
\end{align}
and they generate its commutant in the $q$-oscillator realization of $\mathfrak{o}_{q^{1/2}}(2n)$. We claim that the relations obeyed by these elements are precisely the relations of the higher rank Askey--Wilson algebra $AW(n)$ and that this follows from the Howe duality already observed in Section \ref{sec:comm_AW3}. We shall now explain how this conclusion is reached.

It will be useful to make the following linear transformation in order to work with elements $\Lambda^{A}$ where $A\subseteq[1;n]$ is a set of consecutive indices. 
Form the $\Lambda^{[k,\ell]}$'s as follows:
\begin{align}\label{eq:change_variable}
\begin{aligned}
 \Lambda^{[k;k]}&=\Lambda_{k},\\
 \Lambda^{[k;k+1]}&=\Lambda_{k,k+1},\\
 \Lambda^{[k;k+\ell-1]}&=\sum_{1\leq i<j\leq \ell}q^{i+j-(\ell+1)}\Lambda_{k-1+i,k-1+j}-[\ell-2]_{q^{1/2}}\sum_{i=1}^{\ell}q^{i-\frac{\ell+1}{2}}\Lambda_{k-1+i},\qqquad \ell\geq3,
\end{aligned}
\end{align}
with $\Lambda^{\O}=0$ by convention. Note that there are still $\binom{n+1}{2}$ elements of the form $\Lambda^{[i;j]}$ since $\big|\{[i;j]~|~[i;j]\subseteq[1;n]\}\big|=\binom{n+1}{2}$.

For the sake of comprehensiveness, let us also give here the inverse change of basis:
\begin{align}
\begin{aligned}
 \Lambda_{i}&=\Lambda^{[i;i]},\\
 \Lambda_{ij}&=\Lambda^{[i;j]}+q^{-1}\Lambda^{[i;i]}+\Lambda^{[i+1;j-1]}+q\Lambda^{[j;j]}-q^{-1}\Lambda^{[i;j-1]}-q\Lambda^{[i+1;j]}.
\end{aligned}
\end{align}
We can now make use of the Howe duality observed in Section \ref{sec:comm_AW3}. Recall that the decomposition \eqref{eq:pairing_modules} implied that the quadratic Casimirs of $\mathfrak{o}_{q^{1/2}}(2m)$ were affinely related to intermediate Casimirs of $U_q(\su(1,1))$ embeddings in $(2m)$ copies of itself. This still holds here.

Take $2n$ $q$-oscillators and couple them pairwise, with each couple labelled by $\ol{\imath}\equiv(2i-1,2i)$ in order to obtain an embedding of $U_q(\mathfrak{su}(1,1))^{\otimes n}$ in $U_q(\mathfrak{su}(1,1))^{\otimes2n}$. Nested embeddings then give rise to all the intermediate Casimirs needed to generate the $AW(n)$ algebra. This can be visualized as follows:
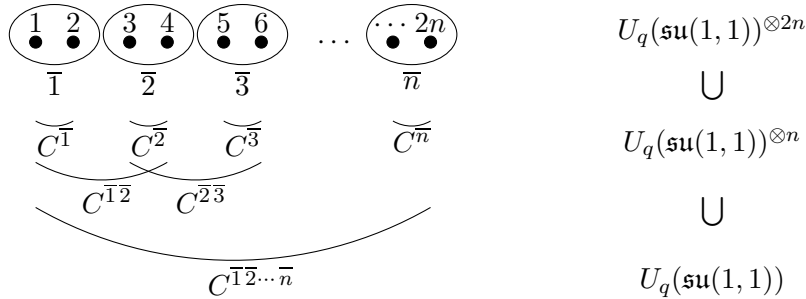
\begin{center}
\begin{tikzpicture}
\draw[fill] (-.25,0) circle [radius=0.08];
\draw[fill] (.25,0) circle [radius=0.08];
\node [above, black] at (-.25,.0) {$1$};
\node [above, black] at (.25,.0) {$2$};
\node [below, black] at (0,-.25) {$\ol{1}$};
\draw[postaction={decorate,decoration={markings,
      mark=at position 0.8 with {\node (a1) {};}}}]
      (0,.1) ellipse (0.6 and 0.4);

\draw[fill] (1,0) circle [radius=0.08];
\draw[fill] (1.5,0) circle [radius=0.08];
\node [above, black] at (1,.0) {$3$};
\node [above, black] at (1.5,.0) {$4$};
\node [below, black] at (1.25,-.25) {$\ol{2}$};
\draw[postaction={decorate,decoration={markings,
      mark=at position 0.8 with {\node (a2) {};}}}]
      (1.25,.1) ellipse (0.6 and 0.4);

\draw[fill] (2.25,0) circle [radius=0.08];
\draw[fill] (2.75,0) circle [radius=0.08];
\node [above, black] at (2.25,.0) {$5$};
\node [above, black] at (2.75,.0) {$6$};
\node [below, black] at (2.5,-.25) {$\ol{3}$};
\draw[postaction={decorate,decoration={markings,
      mark=at position 0.8 with {\node (a3) {};}}}]
      (2.5,.1) ellipse (0.6 and 0.4);
\node [black] at (3.75,.0) {$\cdots$};

\draw[fill] (4.5,0) circle [radius=0.08];
\draw[fill] (5.,0) circle [radius=0.08];
\node [above, black] at (4.5,.0) {$\cdots$};
\node [above, black] at (5.0,.0) {$2n$};
\node [below, black] at (4.75,-.25) {$\ol{n}$};
\draw[postaction={decorate,decoration={markings,
      mark=at position 0.8 with {\node (a3) {};}}}]
      (4.75,.1) ellipse (0.6 and 0.4);

\draw(-.25,-1.05) arc (240:300:.5);
\node [below, black] at (0.,-1.) {$C^{\ol{1}}$};
\draw(1,-1.05) arc (240:300:.5);
\node [below, black] at (1.25,-1.) {$C^{\ol{2}}$};
\draw(2.25,-1.05) arc (240:300:.5);
\node [below, black] at (2.5,-1.) {$C^{\ol{3}}$};
\draw(4.5,-1.05) arc (240:300:.5);
\node [below, black] at (4.75,-1.) {$C^{\ol{n}}$};
\draw(-.25,-1.6) arc (240:300:1.75);
\draw(1,-1.6) arc (240:300:1.75);
\node [below, black] at (0.7,-1.75) {$C^{\ol{1}\,\ol{2}}$};
\node [below, black] at (1.95,-1.75) {$C^{\ol{2}\,\ol{3}}$};
\draw(-.25,-2.2) arc (240:300:5.25);
\node [below, black] at (2.6125,-2.85) {$C^{\ol{1}\,\ol{2}\cdots\,\ol{n\vphantom{1}}}$};
\node [below, black] at (8.75,.5) {$U_q(\su(1,1))^{\otimes2n}$};
\node [below, black] at (8.75,-.25) {$\bigcup$};
\node [below, black] at (8.75,-1) {$U_q(\su(1,1))^{\otimes n}$};
\node [below, black] at (8.75,-1.925) {$\bigcup$};
\node [below, black] at (8.75,-2.85) {$U_q(\su(1,1))$};
\end{tikzpicture}
\end{center}
It remains to give the explicit correspondance between the paired Casimirs $C^{\ol{\imath}\,\cdots\,\ol{\jmath}}$ and the $\Lambda^{[i;j]}$'s (which will be the equivalent of \eqref{eq:duality_casimirs_3}). We already know that they are affinely related, so we start by writing
\begin{align}\label{eq:key_element}
 C^{\ol{1}\,\ol{2}\,\cdots\,\ol{m\phantom{1}\!\!\!}}=\Delta^{(2m-1)}(C)=\frac{1}{\beta_{2m}}\left(\Lambda^{[1;m]}+\alpha_{2m}\right).
\end{align}
After a quick look at the coproduct and the form of the $\Lambda^{[1;m]}$'s one convinces oneself that $\beta_{2m}$ is constant, more precisely $\beta_{2m}=(1+q)^{2}$. It remains to evaluate the $\alpha_{2m}$'s.

The $\alpha_{2m}$ can be obtained by acting with \eqref{eq:key_element} on the ground state $\ket{\vec{0}\,}$ of $2m$ $q$-oscillators and recalling the action of the generators on the realization \eqref{eq:AqN_actions}, that is $\mathscr{L}_{ij}^{\pm}\ket{\vec{0}\,}=\mathscr{J}_{-}^{i}\ket{\vec{0}\,}=0$ and $\mathscr{J}_{0}^{i}\ket{\vec{0}\,}=\frac{1}{4}\ket{\vec{0}\,}$:
\begin{align}
\begin{aligned}
 \left(\Lambda^{[1;m]}+\alpha_{2m}\right)\ket{\vec{0}\,}&=(1+q)^{2}C^{\ol{1}\,\ol{2}\,\cdots\,\ol{m\phantom{1}\!\!\!}}\ket{\vec{0}\,}\\
 \alpha_{2m}\ket{\vec{0}\,}&=(1+q)^{2}\Delta^{(2m-1)}\left(q^{-2\mathscr{J}_0+1} \mathscr{J}_+\mathscr{J}_- + (\mathscr{J}_0)_{q^{2}}(1-\mathscr{J}_0)_{q^{2}}\right)\ket{\vec{0}\,}\\
 &=(1+q)^{2}\left( ({\textstyle\sum_{i=1}^{2m}}\mathscr{J}_0^{i})_{q^{2}}(1-{\textstyle\sum_{i=1}^{2m}}\mathscr{J}_0^{i})_{q^{2}} \right)\ket{\vec{0}\,}\\
 &=(1+q)^{2}\left( (\nicefrac{m}{2})_{q^{2}}(1-\nicefrac{m}{2})_{q^{2}} \right)\ket{\vec{0}\,}\\
 &=-[m]_{q^{1/2}}[m-2]_{q^{1/2}}\ket{\vec{0}\,}
\end{aligned}
\end{align}
We conclude that
\begin{align}\label{eq:complete_affine_1}
 C^{\ol{1}\,\ol{2}\,\cdots\,\ol{m\phantom{1}\!\!\!}}=\frac{1}{(1+q)^{2}}\left(\Lambda^{[1;m]}-[m]_{q^{1/2}}[m-2]_{q^{1/2}}\right),
\end{align}
which finally leads to the desired generalization of \eqref{eq:duality_casimirs_3}:
\begin{align}\label{eq:duality_casimirs_n}
 C^{\,\ol{\imath}\,\cdots\,\ol{j\phantom{1}\!\!\!}}=\frac{1}{(1+q)^{2}}\left(\Lambda^{[i;j]}-[j-i+1]_{q^{1/2}}[j-i-1]_{q^{1/2}}\right).
\end{align}
Upon shifting the Casimirs $C^{\,\ol{\imath}\,\cdots\,\ol{j\phantom{1}\!\!\!}}$ using the procedure \eqref{eq:realiz_AW3_cas}, one finally obtains the desired generating set for $AW(n)$
\begin{align}\label{eq:gen_set_A}
 C_{A}=[2]_{q}-q(q-q^{-1})^{2}C^{\ol{A}}.
\end{align}
By virtue of the affine correspondance between the intermediate Casimirs of $U_q(\su(1,1))$ and the $\os_{q^{1/2}}(2m)$ quadratic Casimirs given in \eqref{eq:duality_casimirs_n}, the $AW(n)$ algebra generated by all $C_A$'s therefore admits two dual descriptions.
\begin{rmk}
The $q\to1$ limit of the expression for the pairing of the Casimirs in \eqref{eq:duality_casimirs_n} coincides with the result obtained for the higher rank Racah algebra $R(n)$. Indeed, for $|A|=2m$, one has
\begin{align}
 C^{A}=-(J_0^{A})^{2}+J_+^{A}J_-^{A}+J_0^{A}=\frac{1}{4}\left(\sum_{\mu<\nu\in A}{L_{\mu\nu}}^{2}-\frac{|A|(|A|-4)}{4}\right).
\end{align}
\end{rmk}

\section{Conclusion}\label{sec:conclusion}
To sum up, we have used the Howe duality to provide two dual pictures of the Askey--Wilson algebra $AW(n)$. In addition to the description of $AW(n)$ as the commutant of $U_q(\mathfrak{su}(1,1))$ in $U_q(\mathfrak{su}(1,1))^{\otimes n}$, we have also depicted the algebra as the commutant of $\mathfrak{o}_{q^{1/2}}(2)^{\oplus n}$ in $q$-oscillator representations of $\mathfrak{o}_{q^{1/2}}(2n)$. We have explained how the multiplicity-free decomposition of the modules of the joint action of the dual pair $\big(U_q(\mathfrak{su}(1,1)),\mathfrak{o}_{q^{1/2}}(2n)\big)$ given in \eqref{eq:pairing_modules} translates to an affine correspondance between Casimirs of $U_q(\mathfrak{su}(1,1))$ and $\mathfrak{o}_{q^{1/2}}(2m)$. This fact was then stressed to be the hallmark of the duality between the two pictures.

The $q\to1$ limit is easily seen to give back results we have previoulsy obtained on the higher rank Racah algebra $R(n)$. Note that in \cite{Gaboriaud2018a} we carried out the dimensional reduction corresponding to the imposition of the $\os(2)^{\oplus n}$ invariance on the oscillator model and this had led us to the generic superintegrable model on the $(n-1)$-sphere \cite{Kalnins2007, Iliev2017}. Such a dimensional reduction has not been performed here as the right $q$-analogues of polar coordinates are not known, but it would be an interesting question to examine in the future.

Another interesting limit is $q\to-1$. This limits yields the higher rank Bannai-Ito algebra $BI(n)$ if one starts from $AW(n)$. In \cite{Gaboriaud2018b} two dual pictures of $BI(n)$ were presented based on a Dirac model. An especially striking result is that the non-naive embeddings of $\osp(1|2)$ associated to non-consecutive tensor product spaces could be explained in the context of the Dirac model by looking at the construction procedure of the higher dimension gamma matrices. It is still an open question to obtain an analogous explanation for $AW(n)$ and the corresponding $U_q(\su(1,1))$ embeddings. We hope to return to this question soon.

\subsection*{Acknowledgments}
The authors would like to thank Nicolas Cramp{\'e}, Hendrik De Bie and Sarah Post for stimulating discussions.
LV wishes to acknowledge the hospitality of the CNRS and of the LAPTh in Annecy where part of this work was done. 
ER and LF are also thankful to the Centre de Recherches Math{\'e}matiques (CRM) for supporting their visits to Montreal in the course of this investigation. 
JG holds an Alexander-Graham-Bell scholarship from the Natural Science and Engineering Research Council (NSERC) of Canada. 
The research of LV is supported in part by a Discovery Grant from NSERC.

\bibliographystyle{unsrtinurl} 
\bibliography{citAW.bib} 
\end{document}